\documentstyle [12pt,epsf] {article}
\begin{document}
\begin{titlepage}
\begin{center}
\vspace{2cm}
\LARGE
Chemical Enrichment and the Origin of the Colour-Magnitude Relation of
Elliptical Galaxies
in a Hierarchical Merger Model\\
\vspace{1cm}
\large
Guinevere Kauffmann$^{1}$ \& St\'{e}phane Charlot$^{2}$  \\
\vspace{0.5cm}
\small
{\em $^1$Max-Planck Institut f\"{u}r Astrophysik, D-85740 Garching, Germany} \\
{\em $^2$ Institut d'Astrophysique du CNRS, 98 bis Boulevard Arago, Paris,
France} \\
\vspace{0.8cm}
\end{center}
\normalsize
\begin {abstract}
In this paper, we present a model of the formation and chemical enrichment of
elliptical galaxies that
differs from the conventional picture in two fundamental ways:
\begin {enumerate}
\item  Ellipticals do not form in a single
monolithic collapse and burst of star formation at high redshift. Instead, most
of their stars
form at modest rates in disk galaxies, which then merge to form the
ellipticals.
\item Galaxies do not undergo ``closed-box'' chemical evolution. Instead,
metals can be
transferred between the stars, cold gas and the hot gas halos of the galaxies.
It is assumed that metals are ejected out of disk galaxies during supernova
explosions
and these metals enter the hot gas component. The fact that
metals are more easily ejected from small galaxies leads to the establishment
of a
mass-metallicity relation for the disk systems. Big ellipticals are more metal
rich because they
are formed from the mergers of bigger disks.
\end {enumerate}
We use semi-analytic techniques to follow the formation, evolution, and
chemical enrichment
of cluster elliptical galaxies in a merging hierarchy of dark matter halos.
The inclusion of the new metallicity-dependent spectral
synthesis models of Bruzual \& Charlot enable us to compute the colours, line
indices and
mass-to-light ratios of these galaxies. We find that with physically realistic
parameters and
with the assumption that feedback is efficient, even in massive
galaxies, we are able to reproduce the slope and scatter of the
colour-magnitude
and the Mg$_2 - \sigma$ relations.  We are not able to match the
increase in $M/L$ for bright ellipticals if we assume their mass to
measure purely the total quantity of stars in the galaxy.
We also study the evolution of these
relations to high redshift. We show that the luminosity-metallicity relation
does not change
with redshift,  but the mean stellar age of the galaxies scales with the
age of the Universe. This is why the evolution of cluster ellipticals  appears
to be well described by simple passive evolution. Finally, we study the
enrichment history of
the intracluster gas. Our models predict that more than 80\% of the metals were
ejected by
galaxies with circular velocities less than 250 km s$^{-1}$ at redshifts
greater than 1.
The metallicity of the ICM
is thus predicted to evolve very little out to $z > 1$.

\end {abstract}
\vspace {0.8 cm}
\end {titlepage}

\section {Introduction}
The relationship between the colours, line strengths and luminosities of
elliptical galaxies
is potentially a source of information about physical processes
important during their formation. The conventional interpretation
of the colour-magnitude relation of ellipticals is that it is primarily a
metallicity
effect (Faber 1977; Dressler 1984; Vader 1986). It is often  assumed that
elliptical galaxies
form monolithically in a single giant burst of star formation at high redshift.
Supernova
explosions heat the gas, and when its thermal energy exceeds its
binding energy, a galactic wind is produced, which is assumed to interrupt
further
star formation and enrichment. Since the binding energy per unit mass of gas is
higher in more massive galaxies, these are able to retain their gas for much
longer times
and so to reach higher metallicities than less massive galaxies (Larson 1974;
Matteuci \& Tornamb\'{e} 1987;
Arimoto \& Yoshii 1987; Bressan et al 1996).
This interpretation is not unique, however, because colours and line strengths
are degenerate
in age and metallicity. As a result, the colour-magnitude relation of local
ellipticals
can also be interpreted as an increase in mean stellar age with luminosity
(Worthey 1996).
In recent work, Kodama \& Arimoto (1997) have pointed out that the evolution of
the colour-magnitude
relation to high redshift does not support the hypothesis that age effects
alone
are responsible for the redder colours of the brighter ellipticals. Following
this,
Charlot, Ferreras \& Silk (1997, in preparation) have explored models where
both
age and metallicity are important, and have shown that several of
these can provide equally good fits to the data.

An alternative approach to interpreting the spectra of ellipticals
is to begin with a more realistic dynamical picture for the formation of these
objects, and
then to study the properties of their stellar populations by applying
spectral synthesis techniques.
Recently, there has been considerable progress in modelling the observed
properties of
ellipticals in a scenario where they are formed by the merging of disk galaxies
in a
universe where structure is built through hierarchical clustering (Kauffmann
1996b,
Baugh, Cole \& Frenk 1996). This work has, for the first time, placed the
theory of the formation and
evolution of elliptical galaxies in a proper cosmological context. In these
models,
star formation takes place mainly in centrifugally-supported disks, formed when
gas
cools and condenses at the centres of virialized halos of dark matter.
A self-regulated balance between cooling and feedback from supernova explosions
maintains star formation at a roughly constant rate in these disks. If two disk
galaxies
merge once their surrounding halos have coalesced, an elliptical galaxy is
formed and all
cold gas present in the merger remnant is transformed instantaneously into
stars in a
``burst''. Kauffmann (1996b) showed that although ellipticals are predicted to
have been assembled
by relatively recent mergers ($z < 1$) in a standard cold dark matter (CDM)
cosmology,
most of their stars form at considerably higher redshift in their
disk-galaxy progenitors. The $V$-band luminosity-weighted stellar age of
cluster ellipticals
was shown to vary between 8 and 12 Gyears and the scatter in the predicted
$V-K$ and
$U-V$ colours was of order 0.04, consistent with the small observed scatter of
Virgo
and Coma cluster elliptical colours (Bower, Lucey \& Ellis 1992). The model
presented in
that paper did not include any scheme for the chemical enrichment of galaxies
and it
was  unable to reproduce the {\em slope} of the observed colour-magnitude
relation.
An important conclusion of this work  was that age effects in hierarchical
models
could not alone account for the redder colours of the more massive ellipticals.

We have now included a simple prescription for chemical evolution in our
semi-analytic
galaxy formation scheme in order to analyze the effect of variations in
metallicity
on the observed properties of ellipticals. Supernovae take place in stars
forming disk galaxies and the explosions cause cold gas and metals to be
ejected into the halo,
where the metals are mixed with the hot halo gas.
The efficiency with which this process occurs depends on
the potential well depth, so that less massive galaxies eject more material
and hence produce fewer stars and metals per unit mass of available cold gas.
The metallicity of disk galaxies thus increases as a function of their mass.
We then show that in a hierarchical merger picture, bright ellipticals form
from mergers of substantially
more massive disks than faint ellipticals. This lays the theoretical groundwork
for
understanding the origin of a mass-metallicity relation for elliptical
galaxies.

We then compare the predictions of the model with observational data. To do
this, we make use
of the new stellar population synthesis models of Bruzual \& Charlot (1997, in
preparation);
these enable us to compute the colours and spectral line indices of evolving
stellar populations
of different metallicities. We show that with physically reasonable choices for
the free
parameters in our model, we are able to reproduce both the slope and the
scatter of the
$U-V$ and $V-K$ colour-magnitude relations of cluster ellipticals. In addition,
we obtain a good
fit to the slope of observed relation between the Mg$_2$ index
and the stellar velocity dispersion $\sigma$, where it is assumed that $\sigma$
simply scales in proportion
to the circular velocity of the halo in which the elliptical galaxy formed.
We also examine the evolution of these relations to high redshift
and show that out to a redshift of 1 in a standard CDM cosmology,
the slopes change very little and  the colours and index values at fixed
luminosity
shift to lower values as
expected for a passively evolving population of stars. At redshifts greater
than one, there is a
slight flattening in the slopes of the relations, caused by the slightly
younger mean ages
of the bright ellipticals relative to the faint ones.

Although the colour-magnitude relations are reproduced very well by our model,
we cannot account for the
observed increase in the mass-to-light ratios of bright elliptical galaxies if
we assume the mass, estimated
using the central line-of-sight velocity dispersion $\sigma$, is purely a
measure of the total quantity
of stars in the galaxy. We hypothesize that since massive ellipticals are built
up through a greater
number of mergers than their less massive counterparts, more dark matter may
end up being mixed with
their stars.
Finally, since our chemical
evolution model tracks the recycling of heavy elements between stars, cold gas,
and the hot
intergalactic medium, we are able to study the enrichment history of the hot
gas in clusters.
With a Sn Ia + Sn II yield of twice solar, we are able to reproduce the
observed ratio of iron mass
to total blue luminosity in nearby clusters (Renzini et al 1993). We show that
80\% of the metals
are produced at redshifts greater than 1 in galaxies with circular velocities
less than 250
km s$^{-1}$. The metallicity of rich clusters is predicted to evolve very
little out to redshifts
of 1 and beyond.

\section {Modelling the Formation and Chemical Enrichment of Elliptical
Galaxies}

\subsection {Semi-analytic models of elliptical galaxy formation in a
hierarchical Universe}

The semi-analytic model we employ is the same as that used by Kauffmann (1996b)
to compute
the stellar ages of elliptical galaxies and bulges formed by mergers in a cold
dark matter
universe. The reader is referred to that paper for more background.  The new
material in the present
paper is a prescription for chemical enrichment and the transfer of metals
between
stars, cold galactic gas and hot halo gas. We describe this
in detail in the next section.  More details
about semi-analytic techniques may be found in Kauffmann \& White (1993),
Kauffmann, White \& Guiderdoni (1993, KWG) , Cole et al (1994) and Baugh, Cole
\& Frenk (1996).

To summarize:\\
An algorithm based on an extension of the Press-Schechter theory due to Bower
(1991) and
to Bond et al (1991) is used to generate Monte
Carlo realizations of the merging paths of dark matter halos from high
redshift until the present.  This algorithm allows all the progenitors of a
present-day object,
such as a cluster, to be traced back to arbitrarily early times.
Dark matter halos are modelled as truncated isothermal spheres and
it is assumed that as the halo forms, the gas relaxes to a distribution that
exactly
parallels that of the dark matter.
Gas then cools and condenses onto a {\em central galaxy} at the core of each
halo. Star formation and feedback processes take place as described in KWG (see
also below).
In practice, star formation in central galaxies takes place at a roughly
constant rate of a few
solar masses per year, in agreement with the rates derived by Kennicutt (1983)
for normal spiral
galaxies.

As time proceeds, a halo will
merge with a number of others, forming a new halo of larger
mass. All gas which has not already cooled is assumed to be shock
heated to the virial temperature of this new halo. This hot gas then
cools onto the central galaxy of the new halo, which is identified
with the central galaxy of its {\em largest progenitor}. The central
galaxies of the other progenitors become {\em satellite galaxies},
which are able to merge with the central galaxy on a dynamical friction
timescale. If  a merger takes place between two galaxies of roughly comparable
mass, the merger remnant is labelled as an ``elliptical'' and all cold gas
is transformed instantaneously into stars in a ``starburst''.
Note that the infall of new gas onto
satellite galaxies is not allowed, and star
formation will continue in such objects only until their existing cold gas
reservoirs are exhausted. Thus the epoch at which a galaxy is accreted by a
larger
halo delineates the transition between active star formation in the galaxy
and passive evolution of its stellar population. The stellar populations of
elliptical merger remnants in clusters hence redden as their stellar
populations age.
Central galaxy merger remnants in the ``field'' are able to accrete new gas in
the form of a disk
to form a ``spiral'' galaxy consisting of both a spheroidal bulge and a disk
component. Note that as in KWG, we suppress the formation of visible stars in
cooling flows
within halos with $V_c >$ 500 km s$^{-1}$ in order to avoid producing central
cluster galaxies
that are too bright and too blue.

The cosmological initial conditions used in this paper  are a $\sigma_{8}=0.67$
CDM
universe with $\Omega=1$ and
$H_{0}= 50$ km s$^{-1}$ Mpc$^{-1}$. Although this model does not match the COBE
measurements of the
amplitude of the microwave background fluctuations, it has become the de facto
standard for galaxy formation studies, since it provides a reasonable fit to
observed small-scale galaxy clustering.

\subsection {The chemical enrichment of galaxies in a hierarchical Universe}

In simple closed-box models of chemical evolution, all metals produced by stars
are retained
within the galaxy. After all the gas has been converted into stars, the
metallicity of the
galaxy is then simply given by the mass of metals produced per
solar mass of long-lived stars that are formed. In our model, metals are
exchanged between
three components: stars, cold galactic gas, and hot halo gas. The routes
through which
this exchange may take place are the following:

\begin {enumerate}
\item {\em Stars $\longrightarrow$ hot and cold gas} As decribed in KWG, the
star formation rate in disk
galaxies is given by
\begin {equation} \dot {M}_{stars} = \alpha M_{cold} /t_{dyn}, \end {equation}
where $M_{cold}$ is the mass of cold gas in the galaxy, $t_{dyn}= R_{vir}/10
V_{c}$ is a measure
of the dynamical time of the galaxy (this assumes that the galaxy collapses by
a factor 10
from the virial radius of the halo), and $\alpha$ is a free parameter.
We assume that stars form with a standard Scalo (1986) IMF truncated at 0.1 and
100 $M_{\odot}$ and
use the instantaneous recycling approximation (e.g. Tinsley 1980) to compute
the
amount of gas and metals returned to the interstellar medium and halo.
A yield $y$ of heavy elements is produced per solar mass of gas converted into
stars.
A fraction $f$ of this yield is uniformly mixed with the hot gas in the halo,
and
the rest is uniformly mixed with the cold gas in the galaxy. The parameter $f$
is introduced
to allow for the possibility that heavy elements may be ejected directly into
the halo
in galactic fountains or chimneys resulting from the simultaneous explosions of
tens or hundreds of supernovae (see Spitzer 1990 for a review).
The gas fraction returned by evolved stars is estimated
from the population synthesis models to be 25 \% and is roughly independent of
metallicity.
This material is added to the cold gas component of the galaxy.

\item {\em Cold gas $\longrightarrow$ hot gas} In our models, it is postulated
that supernova explosions can
release enough energy to drive cold gas into the halo.
For a standard Scalo IMF, the number of supernovae expected per solar
mass of stars formed is $\eta_{SN} \simeq 5 \times 10^{-3} M_{\odot}^{-1}$. The
kinetic energy of the
ejecta from each supernova is $10^{51}$ ergs. If a fraction $\epsilon$ of this
energy is used
to reheat cold gas to the virial temperature of the halo, the amount of cold
gas lost to the
intergalactic medium in time $\Delta t$ may be estimated from simple energy
balance arguments as
\begin {equation} \Delta M_{reheat} = \epsilon \frac {4}{5} \frac {\dot{M}_*
\eta_{SN} E_{SN}}{V_{c}^{2}} \Delta t
\end {equation}
Here $\epsilon$ is a free parameter controlling the efficiency of the feedback
process.
When cold gas is returned to the halo, it is assumed that its metals are
uniformly mixed with the halo gas.

\item {\em Hot gas $\longrightarrow$ cold gas} Metals present in hot halo gas
may incorporated into a galaxy
when that gas is able to cool and condense at the centres of halos. The cooling
rates in halos
are calculated as described in KWG and are based on the model by White \& Frenk
(1991).
The cooling functions vary according to the
metallicity of the hot gas and are interpolated from Figure 9.9 of Binney \&
Tremaine (1987).
\item {\em Mixing during halo mergers} As structure in the
Universe builds up through hierarchical clustering, dark matter halos merge
and the metals contained in their hot gas components are mixed together.
In this way, metals ejected by  star-forming galaxies in the ``field''
at high redshift may end up contributing to the metals in the gas in rich
clusters at $z=0$.

\item {\em Mixing during galaxy mergers} When two galaxies merge within a
common dark matter halo, the metals contained in the cold gas and stars of both
objects will be
mixed in the merger remnant. We divide galaxy-galaxy merging events into two
different classes:
a) accretion events -- this is when one galaxy is less than a third of the mass
of the other one.
b) major mergers -- the galaxies are within a factor of three in mass.
Major mergers are assumed to lead to the formation of elliptical galaxies. All
the cold gas
contained in the galaxies is turned into stars instantaneously in a ``burst''
and the enrichment during this phase
is assumed to be a closed-box process, i.e. no metals escape from the galaxy.
Accretion events do not induce any extra star formation. The metals present in
the cold gas and
stellar component of the satellite are simply added to the metals in the cold
gas and
stars of the central galaxy.
Note that the factor of three is chosen so that the central galaxies
of halos of circular velocity $\sim$ 200 km s$^{-1}$ will, on average, have
bulge-to-disk ratios consistent
with an Sbc-type spiral galaxy (see KWG). Our rules about star formation and
enrichment
during mergers and accretion events have been chosen for the sake of
simplicity. They do not affect the
results presented in this paper because, as we have mentioned, most stars in
ellipticals do not
form in these phases.

\end {enumerate}

To summarize, the main parameters of our model that control the metallicities
of the stellar, cold
gas and hot gas components in galaxies are the following:
\begin {itemize}
\item $\Omega_b$ the density of the Universe in baryons
\item $\alpha$ the star formation efficiency
\item $\epsilon$ the feedback efficiency
\item $y$ the yield of heavy elements (instantaneous recycling assumed)
\item $f$ the fraction of heavy elements ejected directly into
          the hot halo gas without first being mixed with the cold gas
\end {itemize}

As in previous work (see KWG), we have adopted a ``Milky Way'' normalization,
i.e. we tune the
parameters so that the properties of the central galaxy in a halo of $V_c=$ 220
km s$^{-1}$
match those of our own Galaxy. In practice, the tuning procedure operates as
follows.
We first choose a value of $\Omega_b$. $\alpha$ and $\epsilon$ control the
luminosity
and cold gas mass of the galaxy and are adjusted  to obtain a $B$-band absolute
magnitude
of $\sim -20.5$ and a cold gas mass of $6 \times 10^9 M_{\odot}$. The yield is
chosen
so that the galaxy has a $V$-light weighted mean metallicity of 0.7 solar, a
value consistent
with the mean observed metallicity of stars in the solar neighbourhood (see
Tinsley 1980
and references therein).
It should be noted that in our model, since metals are driven out of a galaxy
by winds
and diluted in the surrounding  gas,  the mean metallicity of the stars in a
galaxy will {\em always}
be lower than the yield $y$.  The higher the feedback efficiency, the higher
the value of $y$ required
to obtain a metallicity of 0.7 solar. The parameter $f$ controls
the total mass of metals contained in the hot halo gas. Since there are no
reliable measurements of
the metallicities of galactic halo gas (indeed, X-ray emission from hot gas
around spiral galaxies has
yet to be detected), we tune $f$ to match the abundance of metals observed in
the intracluster
medium. A compilation of data by Arnaud et al (1992) shows that the Fe
abundance in a sample
of nearby clusters is $\sim 0.3$ solar, with substantial scatter. We have thus
chosen $f$
so that the metallicity of the hot gas in halos of circular velocity  1000 km
s$^{-1}$ matches this value.
Note that the properties of a Milky Way-type disk galaxies were explored in
detail in a paper by
Kauffmann (1996a) using a chemical evolution model very similar to the one
described above.
In that paper, it was demonstrated that the classic G-dwarf problem was solved
by rapid early
enrichment in the low mass objects that later merged to form the Galaxy, and
that
the predicted age-metallicity distribution of the stars
in the Galactic disk agreed rather well with a recent compilation of data on
solar neighbourhood stars.

In Table 1, we list parameter values for two representative models. Model A (or
the strong feedback model) has
$\Omega_b=0.1$. For our adopted Hubble Constant of 50 km s$^{-1}$ Mpc$^{-1}$,
this value lies at the upper end of the permitted range of $\Omega_b$
according to current constraints from models of Big Bang nucleosynthesis (see
Turner 1996
for a review), but at the lower end of measurements the fraction of baryons in
rich clusters (White et al 1993; White \& Fabian 1995).
Model A also has a high supernova feedback efficiency.
The value of the yield (twice solar) is rather high, but still roughly
consistent with theoretical
calculations of the mass of different types of heavy elements ejected during
Type II and
Type Ia supernova explosions assuming a standard Salpeter IMF  (Tsujimoto et al
1995 ).
Model B (or the weak feedback model) has $\Omega_b=0.06$.
It has a considerably smaller feedback efficiency and yield.

\section {The New Spectral Synthesis Models}

We compute the spectrophotometric properties of model galaxies using new
population
synthesis models by Bruzual \& Charlot (1997, in preparation). These span the
range of
metallicities $5\times10^{-3}\leq Z/Z_\odot \leq 5$ and include all phases of
stellar
evolution from the zero-age main sequence to supernova explosions for
progenitors more
massive than $8\,M_{\odot}$, or the end of the white dwarf cooling sequence for
less
massive progenitors. The models are based on recent stellar evolutionary tracks
computed
by Alongi et al. (1993), Bressan et al. (1993), Fagotto et al. (1994a, b, c),
and Girardi
et al. (1996), supplemented with prescriptions for upper-AGB and post-AGB
evolution. The radiative
opacities are taken from Iglesias et al. (1992). In the version used here, we
adopt the
library of synthetic stellar spectra compiled by Lejeune et al. (1997b, in
preparation;
see also 1997a) for all metallicities. This library is based on spectra by
Kurucz (1995,
private communication; see also Kurucz 1992) for the hotter (O-K) stars,
Bessell et al.
(1989, 1991) and Fluks et al. (1994) for M giants, and Allard \& Hauschildt
(1995) for
M dwarfs. The Lejeune et al. spectral library also includes semi-empirical
corrections
for blanketing, a well-known limitation of synthetic spectra (see, for example,
Charlot,
Worthey \& Bressan 1996, and references therein). The resulting model spectra
computed for stellar populations of various ages and metallicities have been
checked
against observed spectra of star clusters and galaxies (Bruzual \& Charlot
1997; Bruzual
et al. 1997).

In addition, the models predict the strengths of 21 stellar absorption
features, including
the Mg$_2$ index near 5000~{\AA} that is often used as a spectral diagnostic in
early-type
galaxies. The predictions are computed using the Worthey et al. (1994) analytic
fitting
functions for index strength as a function of stellar temperature, gravity and
metallicity
that are calibrated empirically on a sample of 460 Galactic stars. This
constitutes the standard
``Lick/IDS'' system, in which each index is defined by a central bandpass
bracketted by two
pseudo-continuum bandpasses. The integrated index of a stellar population is
then computed by
weighting the contributions from individual stars by their level of continuum
(see Worthey et
al. 1994, and Bressan et al. 1996 for more detail). It is worth noting that the
most massive
elliptical galaxies exhibit [Mg/Fe] ratios in excess of that found in the most
metal-rich stars
in the solar neighborhood (by $\sim0.2-0.3$ dex; see Worthey, Faber, \&
Gonzalez 1992). While
this may limit the accuracy of the predicted Mg$_2$ indices of bright
elliptical galaxies,
the recent models of Bressan et al. (1997, in preparation) show convincingly
that an enhancement
in light elements at fixed total metallicity has virtually no effect on the
other spectrophotometric
properties of model stellar populations.

A complete discussion of the differences between the spectrophotometric
predictions of
these models with those in previous studies will be presented in Bruzual \&
Charlot (1997).
The typical discrepancies between the properties of stellar populations of the
same
input age and metallicity that are obtained by
using the spectral synthesis models constructed by different groups
of scientists, have already been illustrated by Charlot,
Worthey \& Bressan (1996). These can reach up to 0.05~mag in rest-frame $B-V$,
0.25~mag in
rest-frame $V-K$ and a 25\% dispersion in the $V$-band mass-to-light ratio.
With these uncertainties in mind, we will concentrate more on understanding the
trends seen
in the observations than on obtaining exact fits to the data.

\section {Theory of the origin of a mass-metallicity relation for ellipticals}

As was discussed previously, the assumption that supernova feedback ejects cold
gas and metals
from galaxies and that the efficiency of this process scales as $ 1/V_c^2$,
where $V_c$ is
the circular velocity of the galaxy, means that more massive and luminous
disk galaxies will be more metal rich. Because of the $1/V_c^2$ scaling, the
metallicity-luminosity relation will saturate at large $V_c$ when the feedback
efficiency
becomes negligible and all gas and metals are retained in the galaxy.
For large $V_c$, galaxies are effectively closed boxes. Thus in our scenario,
metals in
the hot gas component, including the gas in rich clusters, are produced
primarily
by low-mass galaxies since these lose their gas very easily. In clusters, most
of the enrichment takes place
at high redshift before the cluster has assembled (see section 8).

In figure 1, we show the metallicity-luminosity relation for {\em disk}
galaxies at $z=0$ in our model.
We have
plotted the $V$-light weighted mean metallicities of the stars in these disks.
The result for
model A is shown as a solid line and that for B as a dashed line.
As expected the relation rises much more steeply for model A, which has much
more efficient feedback.
In model B, the relation saturates at $V$-absolute magnitudes of around $-20$.
The model A relation is
in better qualitative agreement with the observed metallicity-luminosity
relation of spiral galaxies (measured using
emission lines from the gas in HII regions) since this shows no turnover
(Roberts \& Haynes 1994).

\begin{figure}
\centerline{
\epsfxsize=8cm \epsfbox{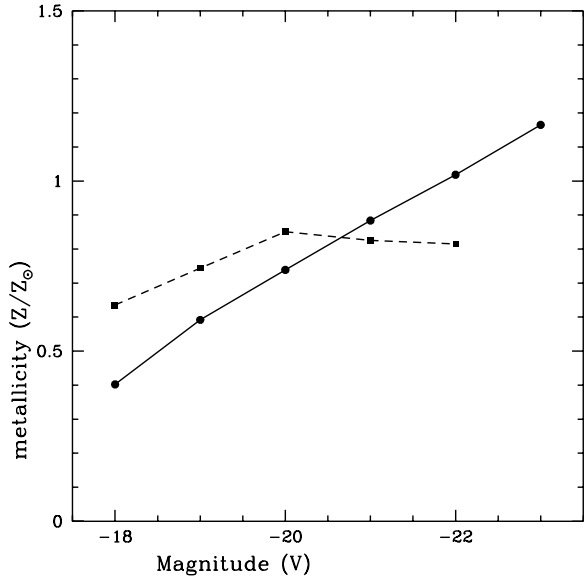}
}
\caption{\label{fig1}
 The metallicity-luminosity relation for disk galaxies in the model:
 the $V$-luminosity weighted mean metallicity of the stars
is plotted as a function of $V$- band absolute magnitude. The solid line is for
the high feedback
 model A and the
dashed line is for the low feedback model B.}
\end {figure}

In the merger model, it is not {\em a priori} clear that ellipticals should
also obey a
metallicity-luminosity relation. For this to be the case, bright ellipticals
have to
be formed from the mergers of {\em systematically bigger disks} than faint
ellipticals.
Figure 2 demonstrates that this is indeed the case. We plot the mean progenitor
mass
of ellipticals in present-day clusters of circular velocity 1000 km s$^{-1}$
versus their stellar mass.
The mean progenitor mass is computed by weighting each star in the galaxy
by the mass of the disk-galaxy progenitor in which it formed and is expressed
in units of the
total mass of the elliptical.
As can be seen, the progenitors of elliptical galaxies of
$ 2 \times 10^{11} M_{\odot}$ are typically
10 times more massive than those of $10^{10} M_{\odot}$ ellipticals.
The error bars show the 5th to 95th percentile spread in mean progenitor mass
among ellipticals in
each mass bin.

\begin{figure}
\centerline{
\epsfxsize=8cm \epsfbox{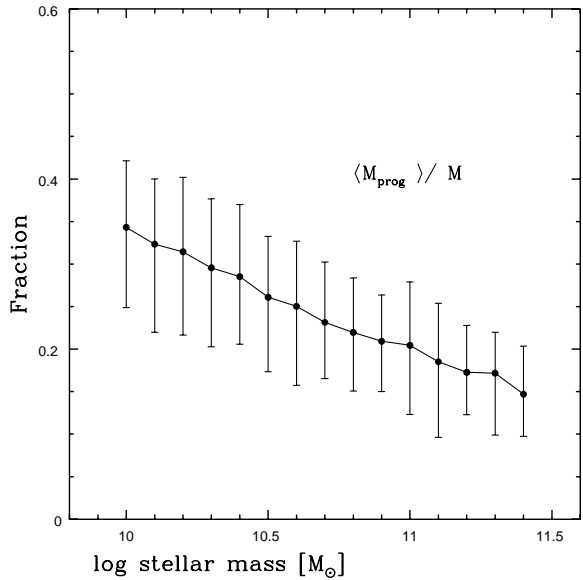}
}
\caption{\label{fig2}
 The mean progenitor mass of ellipticals in clusters of circular velocity
1000 km s$^{-1}$  is plotted as a function
of their stellar mass. The progenitor mass is expressed as a fraction of the
final mass of the elliptical. Errorbars show the 5th-95th percentile range in
progenitor mass.}
\end {figure}

In figure 3 , we show the metallicity-luminosity and age-luminosity relations
for present-day cluster
ellipticals. The left panel is for model A and the right panel is for model B.
We plot $V$-light weighted stellar quantities as before. Note that the ages
differ somewhat from
those given in Kauffmann (1996b), because the $V$-light weighting
has changed as a result of the implementation of the new spectral
synthesis models.
The rms scatter in age is  1.2 Gyear and the scatter in metallicity at fixed
luminosity is
0.14 in solar units.
Model A again has a much steeper metallicity-luminosity relation than model B.
Both models show very little trend in age with luminosity. If anything, the
brighter ellipticals have
somewhat younger ages. We conclude, therefore, that
in the hierarchical merger model, the observed colour-magnitude relation for
elliptical
galaxies must arise as a metallicity effect.

\begin{figure}
\centerline{
\epsfxsize=11cm \epsfbox{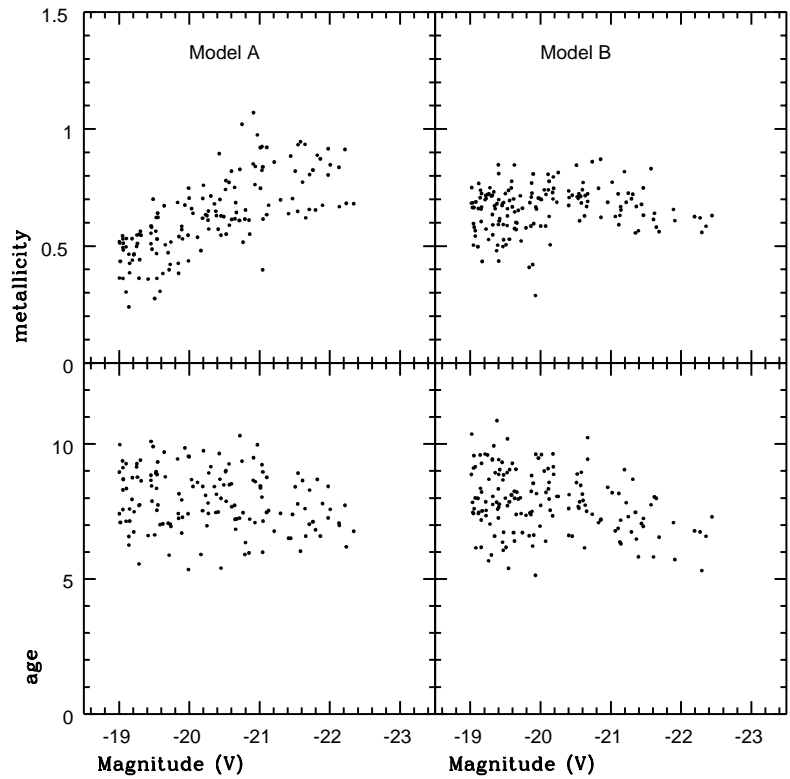}
}
\caption{\label{fig3}
The metallicity-luminosity and age-luminosity relations of elliptical galaxies
in clusters of circular velocity 1000 km s$^{-1}$ at the present day .
$V$-luminosity weighted quantities
are plotted, as before.}
\end {figure}

\section {The colour-magnitude and Mg$_{2}$-$\sigma$ relation for cluster
ellipticals}

In figure 4, we show the $U-V$ and $V-K$ colour-magnitude relations for cluster
ellipticals.
Results for model A are shown in the left-hand panel and results for model B in
the right.
The straight lines are the fits to the $U-V$ and $V-K$ colour-magnitude
relations of elliptical galaxies in
the Virgo and Coma clusters derived by Bower, Lucey \& Ellis (1992, hereafter
BLE). Model A fits the observations
very well. We have computed least-squares estimates of the slopes and scatter
of the Model A
relations and compare them in Table 2 with the values derived by BLE for the
Coma and Virgo ellipticals.
Note that the galaxies plotted in figure 4 are from 5 different Monte Carlo
realizations of the
formation of a cluster of $V_c=1000$ km s$^{-1}$. Even so, the relation has
very small scatter.
Since model A fits the data so much better than model B, in future we show
results only for model A.

\begin{figure}
\centerline{
\epsfxsize=15cm \epsfbox{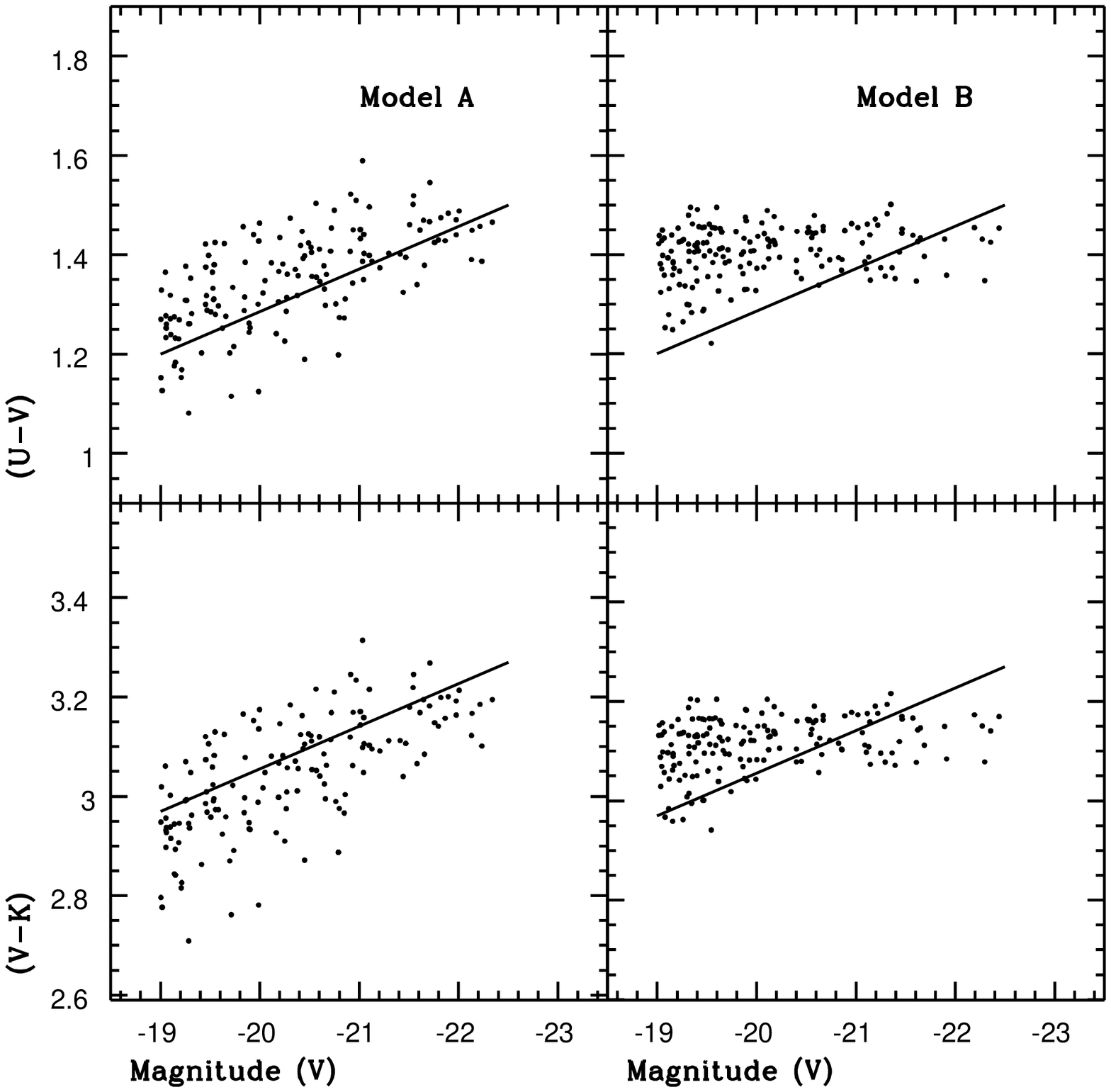}
}
\caption{\label{fig4}
 The $U-V$ and $V-K$ colour-magnitude relations of ellipticals galaxies in
clusters
 of circular velocity 1000 km s$^{-1}$. The solid lines are the fits to the
data on Coma
and Virgo ellipticals taken from Bower, Lucey \& Ellis (1992).}
\end {figure}

It is well known that the Mg$_2$ line index correlates strongly with the
stellar velocity dispersion
of elliptical galaxies (e.g. Bender, Burstein \& Faber 1993).
In our models, this relation is less straightforward to derive as we do not
have a
dynamical model for computing $\sigma$. For simplicity, we have assumed a
constant ratio
$V_c/\sigma$, where $V_c$ is the circular velocity of the halo in which the
elliptical
formed during its last major merger. If the stars were a non-self-gravitating
test population with
density profile $\propto r^{-3}$ within a singular isothermal halo, one would
have
$V_c/\sigma=3^{1/2}$ (Gunn 1982).

The model Mg$_2$ indices, computed using the population synthesis models as
described in section 3, are
{\em global indices}, i.e.
they are averaged over the entire stellar mass of the galaxy. For nearby
ellipticals, however,
the observed indices come from measurements in the central regions of the
galaxies. Since
the abundance of magnesium is known to fall off quite strongly with radius, we
are forced to
correct for this effect if we want to make a meaningful comparison with the
data.
We have used the observed  Mg$_2$-radius relation derived by Davies, Sadler \&
Peletier (1993)
(${\rm Mg}_2 \propto -0.059 \log (R/R_e)$, $R_e$ is the effective radius of the
elliptical), to transform the global
indices to central values. Since our model does not make any prediction for
the sizes of elliptical galaxies, we use the observed relation between
$V$-luminosity and
effective radius given in Guzman, Lucey \& Bower (1993) to assign  a value of
$R_e$ to each of
the model galaxies.

The uncorrected and corrected Mg$_2$ - $\sigma$ relations
are shown in figure 5. The solid line is a fit to the data of J{\o}rgensen,
Franx \& Kjaergaard (1996).
These authors quote Mg$_2$ indices corrected to a circular aperture with
diameter 1.19 h$^{-1}$ kpc,
equivalent to 3.4 arcsec at the distance of the Coma cluster.
Since there is substantial scatter both in the observed Mg$_2$ abundance
gradients
of ellipticals and in the values of $R_e$ at fixed luminosity, which we ignore
in making the
corrections, we will not attempt to analyze the scatter in the relation. What
is clear,
however, is that the slope of the corrected relation matches the observations
rather well.
To fit the observations, we require $V_c/\sigma \sim 2$.

\begin{figure}
\centerline{
\epsfxsize=10cm \epsfbox{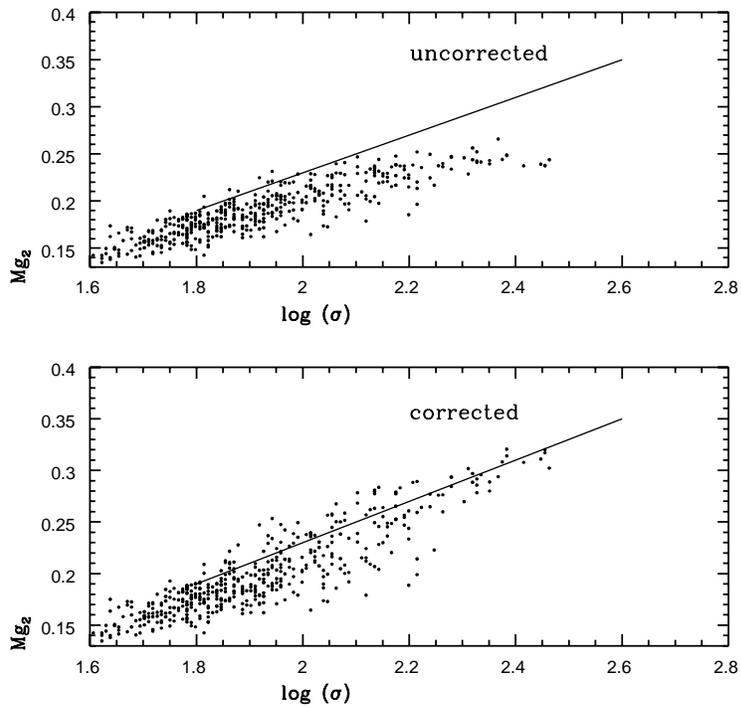}
}
\caption{\label{fig5}
The corrected and uncorrected Mg$_2 - \sigma$ relations computed from our
models. In the lower panel, the Mg$_2$ indices have been transformed from
global
indices to central indices as described in the text. The line is a fit to the
data from
J{\o}rgensen, Franx \& Kjaergaard (1995).}
\end {figure}

In figure 6, we plot the $M_B - \sigma$ Faber-Jackson relation for ellipticals
(also assuming $V_c/\sigma =2$). The line is a fit to the $M_B - \sigma$
relation of Coma
and Virgo cluster ellipticals from Dressler et al (1987). It is reassuring that
we are able to
fit both the $M_B - \sigma$ and the Mg$_2 - \sigma$ relation with the same
parameters. This gives us confidence that the different outputs of the spectral
synthesis models
are giving consistent results.

\begin{figure}
\centerline{
\epsfxsize=8cm \epsfbox{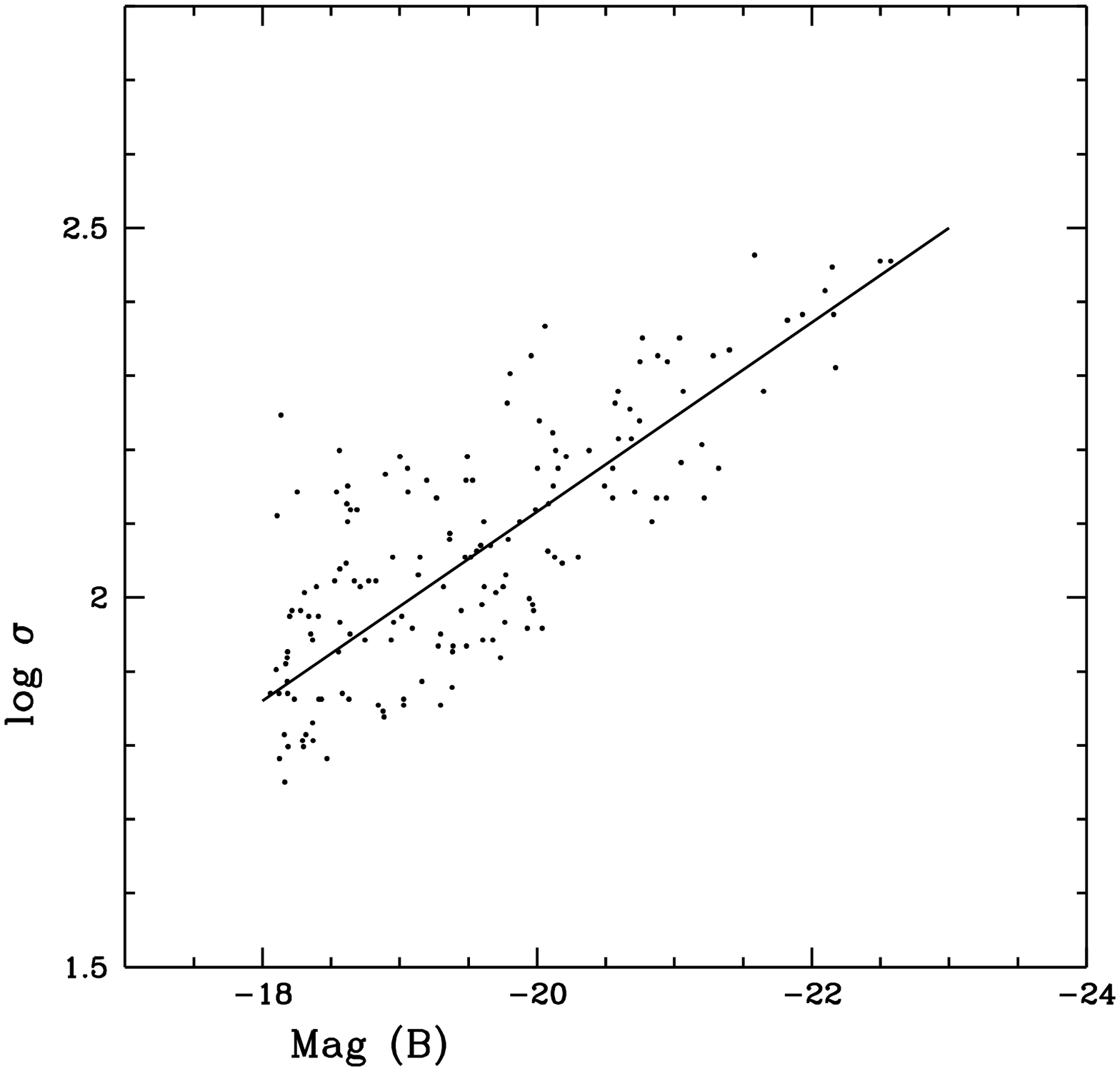}
}
\caption{\label{fig6}
 The  $B$-band ``Faber-Jackson'' relation for ellipticals in the model.
The line is a fit to the data for Coma Cluster ellipticals taken from Dressler
et al (1987).}
\end {figure}

\section {The mass-to-light ratios of cluster ellipticals}

If the luminosity profiles and the dynamical structure of all elliptical
galaxies are similar, the
virial theorem implies that the mass-to-light ratio $M/L$ is a simple function
of the
effective radius $R_e$, the mean surface brightness $\langle I \rangle _e$, and
the
central velocity dispersion $\sigma$. All three quantities may be easily
determined
observationally.

It should be noted that the $M/L$ ratio is a measure of both the stellar
population and the dark
matter content of galaxies. Our models can be used to predict the {\em stellar}
$M/L$
ratios of ellipticals. In figure 7, we show $(M/L)_V$ and
$(M/L)_K$ as a function of stellar mass. The lines are fits to recent data on
Coma cluster ellipticals taken from Mobasher et al (1997). It should be noted
that the
values we obtain for our stellar $M/L$ ratios are very sensitive to the choice
of
IMF. The results shown in figure 6 are for a Scalo (1986) IMF with a lower-mass
cutoff
of 0.1 $M_{\odot}$. A Salpeter IMF with the same lower cutoff gives $M/L$
ratios a factor of 2-3 higher. The slope
of the relation is, however, independent of the IMF.
As can be seen, our models fail to reproduce the observed increase in $M/L$
with mass.
Our $(M/L)_K$ ratios actually decrease with mass, whereas Mobasher et al
find $(M/L)_K \propto M^{\alpha}$, with $\alpha=0.14 \pm 0.01$.

\begin{figure}
\centerline{
\epsfxsize=8cm \epsfbox{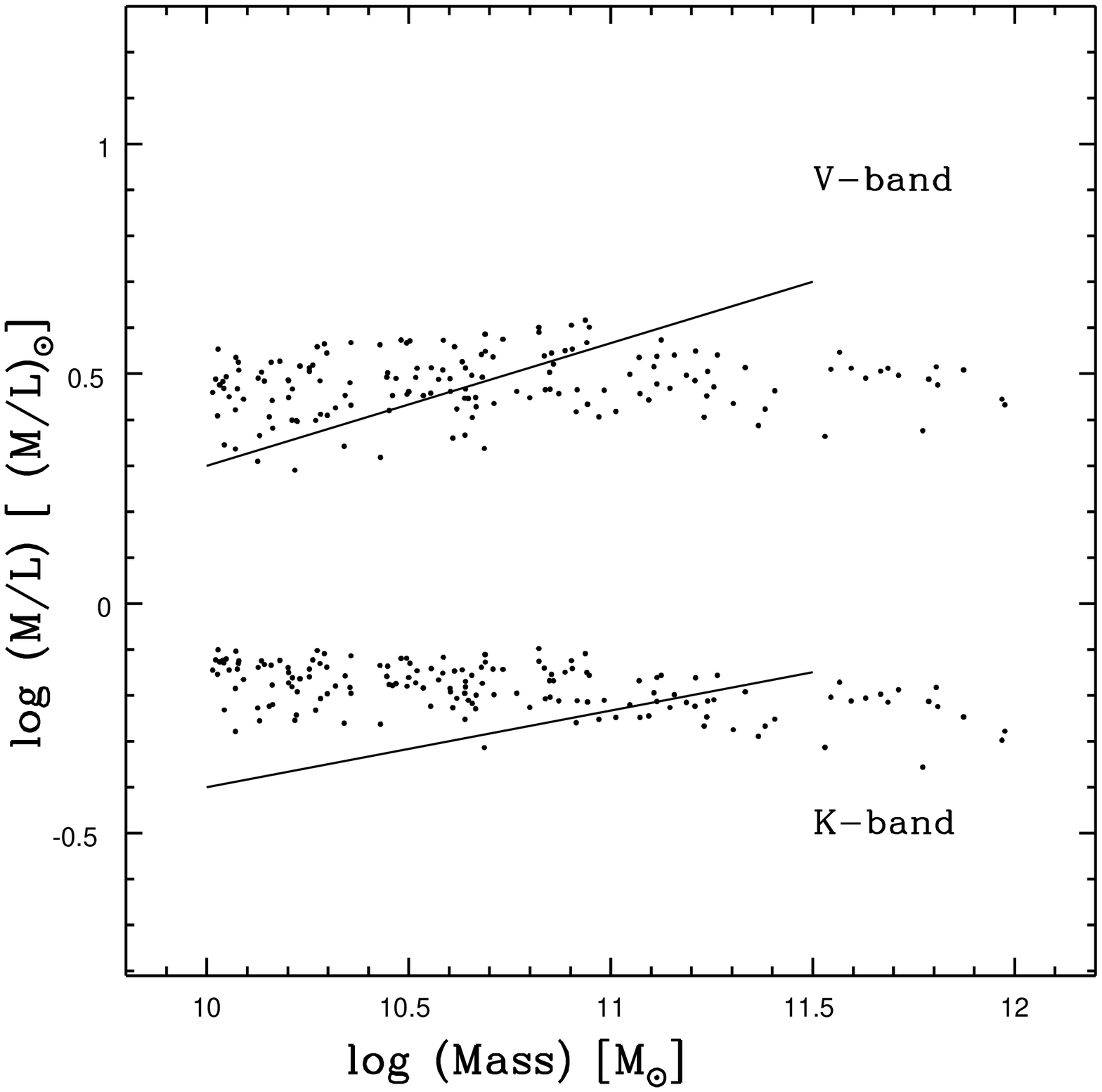}
}
\caption{\label{fig7}
The $V$-band and $K$-band mass-to-light ratios of elliptical galaxies in
clusters
are plotted as a function of their stellar mass.
The lines are fits to data from Mobasher et al (1997).}
\end {figure}

One way of resolving this problem would be for more massive ellipticals to
contain
a larger fraction of dark matter than less massive ellipticals. In our model,
this
requires that more effective mixing of dark matter with stars take place during
the formation
of more massive ellipticals. As can be inferred from figure 2, massive
ellipticals have typically
more merging events than less massive objects, so this appears plausible.
Observationally, evidence for dark halos
around ellipticals has been found through the studies of HI kinematics (e.g.
Franx et al 1994),
X-ray emission (e.g. Forman et al 1994), radial velocities of planetary nebulae
and globular
clusters (e.g. Mould et al 1990, Hui et al 1994), gravitational lensing (e.g.
Maoz \& Rix 1993)
and measurements of the shape of the stellar line-of sight velocity
distribution (e.g. Carollo et al 1995).
More recently, Rix et al (1997) have analyzed the velocity profile of the
elliptical galaxy
NGC 2434 and show that roughly half the mass within an effective radius is
dark.

It should also be noted that recent analyses of the light profiles
of elliptical galaxies show a slight breaking of the generally assumed
homology.
Allowing for this effect can substantially reduce the tilt of the $M/L$
relation
(Hjorth \& Madsen 1996; Graham \& Colless 1997). This would also
bring our results more in line with the data, and the variation in merger
number with
luminosity might offer an explanation for the variation in profile shape.

Finally, Chiosi et al (1997) claim to obtain a good fit to the observed $M/L$
relation using an
IMF that varies as a function
of the temperature, density and velocity dispersion of the gas that forms the
elliptical.
Their model does not attempt to place elliptical formation in its cosmological
context.

\section {Evolution at high redshift}

In figure 8, we show age-luminosity and age-metallicity relations
for elliptical galaxies in clusters of circular velocity 1000 km s$^{-1}$ at
$z=0.4$ and $z=1.5$.
As can be seen, the metallicity-luminosity relation does not evolve very much
with redshift.
By $z=1.5$, it has somewhat larger scatter, but an $M_V=-21$ elliptical still
has a metallicity of 0.6 solar
on average, compared with 0.75 solar at the present day. The age-luminosity
relation shifts downwards with
redshift, and the magnitude of the shift is roughly $t_0- t(z)$, where $t(z)$
is the age of the Universe
at redshift $z$. Note that the scatter in age is {\em smaller} at high
redshift, reflecting the
compression of the time interval over which the stars in the elliptical are
constrained to form.
Thus in our models, elliptical galaxies of a given mass in high redshift
clusters form and are
enriched in much the same
way as elliptical galaxies of the same mass in clusters at the present day.
By selecting ellipticals
in the richest clusters, one is automatically  {\em biasing} ones sample
of galaxies to the objects that turned around, collapsed
and merged at the very highest redshifts.
It is for this reason that the properties of these galaxies
{\em appear} to follow the so-called ``passive evolution'' prediction (e.g.
Aragon-Salamanca et al 1993;
Stanford et al 1995; Dickinson 1995; Franx \& van Dokkum 1996; Bender, Ziegler
\& Bruzual 1996; Ellis et al
1996).
In practice, however,  the ellipticals in the high $z$ clusters are not
necessarily the direct progenitors
of the Es in present-day clusters, so it is incorrect to set constraints on the
formation of the
present-day elliptical
galaxy population using these objects (Charlot \& Silk 1994). This can only be
done using a complete
redshift survey of these objects (Kauffmann, Charlot \& White 1996).

\begin{figure}
\centerline{
\epsfxsize=15cm \epsfysize=11cm \epsfbox{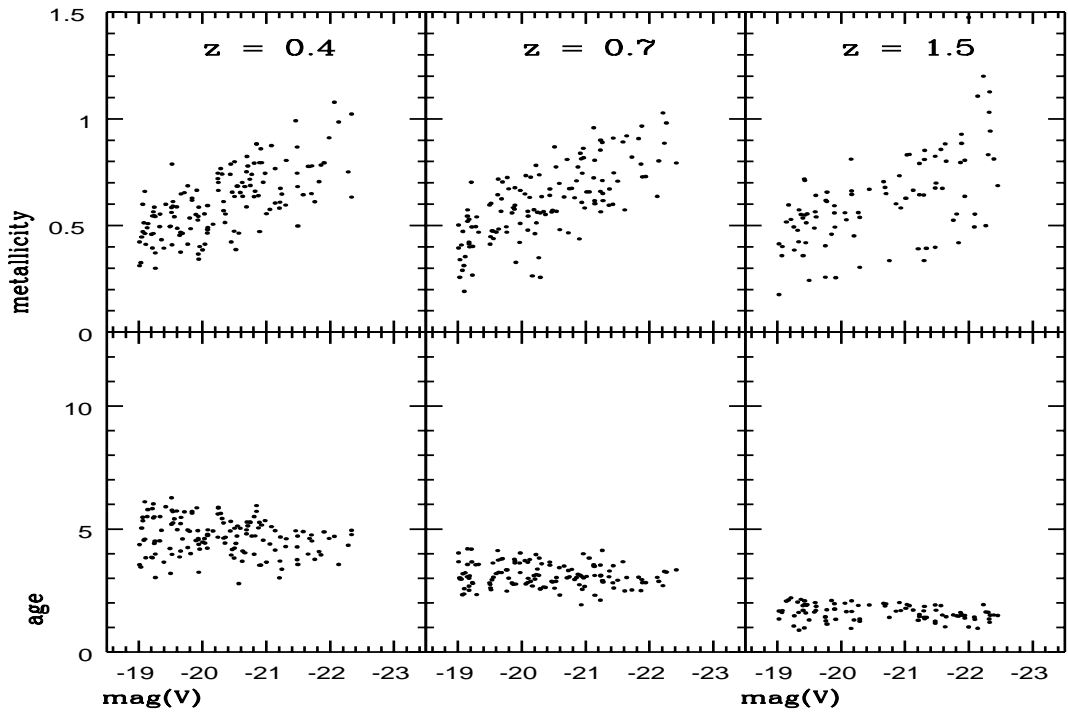}
}
\caption{\label{fig8}
The metallicity-luminosity and age-luminosity relations of elliptical galaxies
in clusters of circular velocity 1000 km s$^{-1}$ at $z=0.4$, $z=0.7$ and
$z=1.5$}
\end {figure}

In figure 9 we show the predicted  restframe $U-V$ colour-magnitude
relation at four different redshifts.
For reference, the solid line is the observed relation of BLE at $z=0$. The
dashed line
indicates the shift in colour expected for a passively evolving population
of solar metallicity with a formation redshift of 10.
Note that there is a progressive flattening in the slope of the
colour-magnitude
relation with redshift, although it only becomes really noticeable at $z > 1$.
This is caused by the fact that the bigger ellipticals have slightly younger
mean ages
in our model, and the effect of this on  observables such as colour and
magnitude, is much
more pronounced at early times. The Mg$_2 - \sigma$ relation
also shows a flattening at $z > 1$, although it is a smaller effect than in the
$U-V$
colours. The slope and scatter
of the rest-frame $U-V$ colour-magnitude relation at each redshift are listed
in Table 2.
The scatter in the rest-frame $U-V$
colours of ellipticals in clusters at $z \sim 0.54$ has been analyzed in detail
by
Ellis et al (1997). They find no change with respect to the $z=0$ relation,
which is
in good agreement with our results.

\begin{figure}
\centerline{
\epsfxsize=15cm \epsfbox{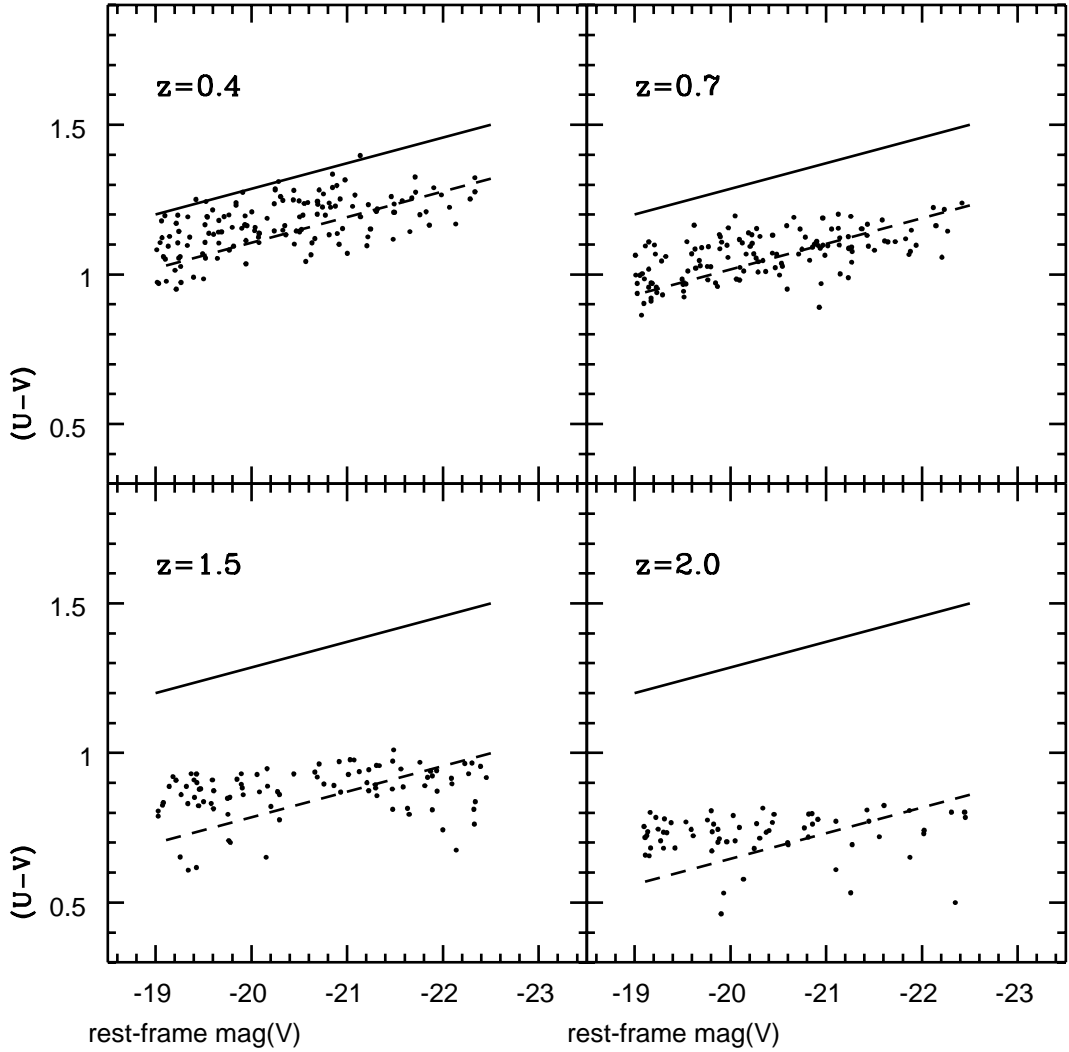}
}
\caption{\label{fig9}
The rest-frame $U-V$  colour-magnitude relations of elliptical galaxies in
clusters
 of circular velocity 1000 km s$^{-1}$ at $z=0.4$, 0.7, 1.5 and 2.0.
The solid lines are the $z=0$ data from Bower, Lucey \& Ellis (1992). The
dashed lines
show the shift in colour expected for a passively evolving population with
$z_{form}=10$. }
\end {figure}

\section {Metals in the intracluster gas}

Since our model follows the exchange of metals between the stars, cold gas and
hot gas component of
galaxies, we are in a position to study the origin of metals in the
intracluster medium. Recall
that we chose the parameter $f$ to obtain a mean metallicity of 0.3 solar in
the hot gas in a halo
with circular velocity 1000 km s$^{-1}$. Rather than  the mass fraction of
metals, many analyses have
focused on the mass-to-light ratio of heavy elements in the ICM, as this
quantity is independent
of the degree to which the metals have been mixed with primordial gas. Most
studies concentrate on
iron, as this is the only element for which it has been possible, until
recently, to derive
accurate abundances using X-ray spectroscopy.  For example, Renzini et al
(1995)
define the iron mass-to-light ratio (IMLR) as the ratio of the iron mass in the
ICM to
the total $B$-band light of the cluster.  They derive values of
$M^{ICM}_{Fe}/L_B$ in the range
0.01-0.02 $M_{\odot}/L_{\odot}$. The clusters we have modelled, which have mass
$5 \times 10^{14}
M_{\odot}$ and a baryon fraction of 10\%, typically contain a total $B$-band
luminosity
of $2 \times 10^{12} L_{\odot}$, of which 60-70\% is in early-type galaxies.
Our  total $B$-luminosity and
E/S0 fraction agree well with the values measured for the Virgo cluster
(Kraan-Korteweg 1981). We obtain
$M^{ICM}_{Fe}/L_B =$ 0.015 $M_{\odot}/L_{\odot}$, in good agreement with the
observations.
Recall that our model has a yield of twice solar. This is larger than the
conventional value
adopted in most analyses (eg Arnaud et al 1992). However, as discussed by
Renzini et al, the uncertainties in
theoretical estimates of the yield of nucleosynthesis products from supernova
explosions are rather large.
Recent calculations by Tsujimoto et al (1995) actually give a (Sn II+Sn Ia) Fe
yield close
to twice solar, assuming a Salpeter IMF and the Sn Ia rate that best reproduces
the abundance
pattern of heavy elements in the solar neighborhood. We note that, according to
Loewenstein \& Mushotsky (1996),
the abundances of elements such as O, Ne, Si and S in nearby clusters may be
even higher than that
of Fe, although the errors in the measurements are still quite large.
The theoretical interpretation of these elements is in principle much simpler
than that of Fe, because
they are produced almost exclusively in Sn II explosions.
Matching these results might require yields even higher than twice solar and
would most likely only be
compatible with a top-heavy IMF. We leave a more detailed
examination of these issues to future work.

We can also use our models to study the history of the enrichment of the ICM
and the
distribution in mass (or circular velocity) of the galaxies that ejected the
metals.
This is illustrated in figure 10, where we plot the fraction of the total
metals present in the
present ICM that was ejected at a given redshift and from a galaxy of given
circular velocity.
As can be seen, more than 80\% of the metals were ejected at redshifts greater
than 1,
60 \% at redshifts greater than 2 and 40\% at redshifts greater than 3.
More than 40 \%  of the ejection was done by galaxies with circular velocities
less than
125 km s$^{-1}$, i.e. ``dwarf'' galaxies have contributed significantly to the
metals in the ICM. Because the enrichment predominantly occurred at high
redshift, our
models predict that abundances in the ICM should evolve very little with
redshift.
Indeed, by $z=1.5$, the model predicts that the ICM  metallicity is still $\sim
0.25$ solar.
Our results are in good agreement with the lack of evolution
seen in the Fe abundances of clusters out to $z \sim 0.3$ as measured by the
ASCA satellite
(Mushotsky \& Loewenstein 1997), although the model abundances show
significantly less scatter
than the data, suggesting that star formation in real cluster galaxies may be
more stochastic
than we have assumed.

\begin{figure}
\centerline{
\epsfxsize=9cm \epsfbox{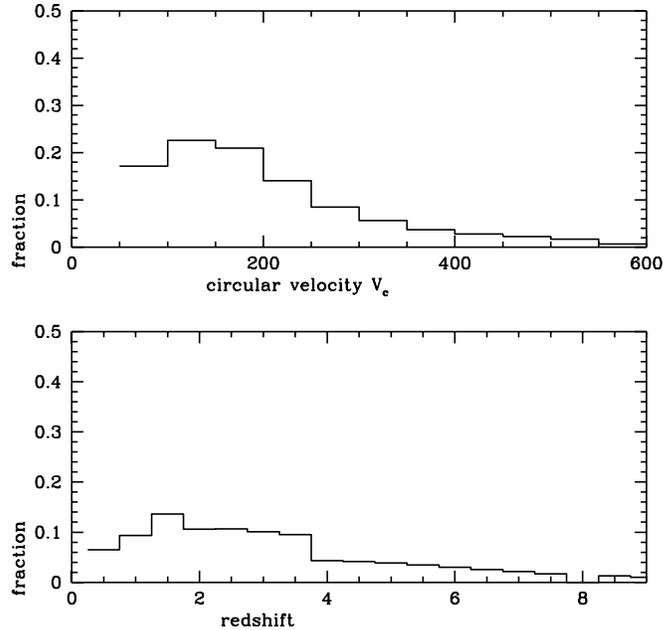}
}
\caption{\label{fig10}
{\em Top panel}: the fraction of metals in the intracluster medium of a
present-day cluster with circular velocity 1000 km s$^{-1}$ ejected by disk
galaxies with circular velocity
$V_c$. {\em Bottom panel}: the fraction of metals in the ICM today that were
ejected at reshift $z$.}
\end {figure}

\section {Discussion}
The failure to match the slope of the colour-magnitude relation of elliptical
galaxies
has been a serious defect of all previous attempts to model the formation of
galaxies
within the framework of a hierarchical cosmology (Lacey et al 1992 ; Kauffmann
1996b; Baugh, Cole \& Frenk 1996).
We have demonstrated that the inclusion of chemical evolution in the model
solves this
problem. With a physically realistic choice of parameters we now obtain good
fits to the $U-V$ and $V-K$ colour-magnitude relations as well as the Mg$_2-
\sigma$
relation. It is interesting that the colour-magnitude relation only extends to
the
brightest ellipticals if we require that feedback be very
efficient,  even in relatively massive galaxies. Using the value of $\epsilon$
given in Table 1 for model A, we can rewrite the cold gas reheating rate in
equation 2 as
\begin {equation} \frac {dM_{reheat}} {dt} = \left ( \frac {280 km s ^{-1}}
{V_c} \right )^2
\frac {dM_*} {dt} \end {equation}
This implies that  the reheating rate is 60\% larger than the star formation
rate for our own Milky Way.
Because of the efficient feedback, our best fit model also requires a high
value of $\Omega_b$, so that
enough gas is able to cool and form the stars of the brightest galaxies.

Note that we have studied only global colours and metallicity indices values in
this paper. It is well-known
that elliptical galaxies exhibit strong gradients in their observed properties.
In particular
the nuclei of ellipticals host stellar populations with  high metallicity
( $\sim 0.5$ to $\sim 3$ Z$_{\odot}$), a magnesium overabundance with
respect to iron, and  a small spread in metallicity.
 As discussed by Greggio (1997), the first two
characteristics of the nuclei can be explained if
gas is processed through multiple
generations of stars on a short timescale. The small metallicity spread
arises if the gas turning into stars
in the nuclei has been pre-enriched. In N-body plus hydrodynamics simulations
of the
merger of two disk galaxies, gas is seen to be
driven very effectively to the centre of the  merger remnant,  where it will
likely
turn into stars on a short timescale (Barnes \& Hernquist 1996). Whether this
picture can account
for the observed properties of elliptical nuclei has yet to be investigated in
detail. In
principle, it would be an
interesting test of our models, which are able to predict when the mergers
occurred, and
the metallicity and mass of the gas in the disk-galaxy progenitors.

As in many other analyses, we  fail to account for the tilt of the $M/L$
relation, if $M$ is purely
a measure of the stellar mass of the galaxies (Djorgovski \& Santiago 1993;
Worthey 1994).
As we have discussed, there are several possible solutions to this problem, but
one
interesting one would be if more massive ellipticals contained a greater
percentage of dark matter in their inner regions than less massive ellipticals.
This might occur if dark matter is more efficiently mixed with the
stars during the formation of the more massive systems. Clearly this is a
problem
that needs to be addressed using N-body simulations that take into account
the full dynamics of the problem.

We have also predicted that the mass-metallicity relation for elliptical
galaxies
in clusters remains virtually unchanged out to high redshift.
It is only the age of the galaxies that ought to decrease.
This is a prediction that might be testable using stellar population indicators
that are able to disentangle age and metallicity effects (eg Jones \& Worthey
1995). In addition, we predict
that there should be a slight flattening in the slope of the colour-magnitude
and
Mg$_2- \sigma$ relations at redshifts greater than 1, reflecting the fact
that more massive ellipticals are slightly younger in age than the less massive
ones.

Finally, X-ray satellites, such as ASCA, are beginning to provide a wealth of
information
about the abundances many different elements in the intracluster gas, as well
as about the evolution
in ICM abundances to high redshift (e.g. Mushotsky \& Loewenstein 1997). This
information
will provide further constraints
on the models, in particular on our prescriptions for star formation and
supernova feedback.
These are indeed the most uncertain elements of our galaxy formation scheme.

\vspace{0.8cm}

\large
{\bf Acknowledgments}\\
\normalsize
We thank Gustavo Bruzual, Marijn Franx, Bodo Ziegler and Alvio Renzini for
helpful discussions.
This work was carried out under the
auspices of EARA, a European Association for Research in Astronomy, and the TMR
Network on Galaxy Formation and Evolution  funded by the European Commission.

\pagebreak

\vspace {1.5cm}
\normalsize
\parindent 7mm
\parskip 8mm

{\bf Table 1:} The parameters for the models
\vspace {0.3cm}

\begin {tabular} {lccccc}
 & $\Omega_b$ & $\alpha$ & $\epsilon$ & $y$ & $f$ \\
 Model A & 0.1 & 0.2 & 0.4 & 2.0 & 0.3 \\
 Model B & 0.06 & 0.1 & 0.05 & 1.2 & 0.45 \\
\end {tabular}

\vspace {3 cm}

{\bf Table 2:} Best-fit regression parameters
\vspace {0.3 cm}

\begin {tabular} {lcc}

& slope & Std. Dev. \\
BLE ($U-V$) & $-0.076$ & 0.037 \\
BLE ($V-K$) & $-0.080$ & 0.048 \\
Model A ($U-V$) $z=0$ & $-0.067$ & 0.019 \\
Model A ($V-K$) $z=0$ & $-0.115$ & 0.035 \\
Model A ($U-V$) $z=0.4$ & $-0.057$ & 0.016 \\
Model A ($U-V$) $z=0.7$ & $-0.052$ & 0.015 \\
Model A ($U-V$) $z=1.5$ & $-0.038$ & 0.017 \\
Model A ($U-V$) $z=2.0$ & $-0.031$ & 0.021 \\

\end {tabular}
\vspace {0.3 cm}

``Standard deviation'' is calculated using the Median  \\
Absolute Difference technique, as in BLE

\pagebreak
\Large
\begin {center} {\bf References} \\
\end {center}
\normalsize
\parindent -7mm
\parskip 3mm

Allard, F., \& Hauschildt, P.H. 1995, ApJ, 445, 433

Alongi, M., Bertelli, G., Bressan, A., Chiosi, C., Fagotto, F., Greggio, L.,
\& Nasi, E. 1993, A\&AS, 97, 851

Arag\'on-Salamanca, A., Ellis, R.S., Couch, W.J. \& Carter, D., 1993, MNRAS,
262, 764

Arimoto, N. \& Yoshii, Y., 1987, A\&A, 173, 23

Arnaud, M., Rothenflug, R., Boulade, O., Vigroux, L. \& Vangioni-Flam, E.,
1992, A\&A, 254, 49

Barnes, J.E. \& Hernquist, L., 1996, ApJ, 471, 115

Baugh, C.M., Cole, S. \& Frenk, C.S., 1996, MNRAS, 283, 1361

Bender, R., Burstein, D., \& Faber, S.M. 1993, ApJ, 411, 153

Bender, R., Ziegler, B. \& Bruzual, B., 1996, ApJ, 463, L51

Bessell, M.S., Brett, J., Scholtz, M., \& Wood, P. 1989, A\&AS, 77, 1

---------. 1991, A\&AS, 89, 335

Binney, J. \& Tremaine, S., 1987, Galactic Dynamics, Princeton University Press

Bond, J.R., Cole, S., Efstathiou, G. \& Kaiser, N. 1991, ApJ, 379, 440

Bower, R. 1991, MNRAS, 248, 332

Bower, R., Lucey, J.R. \& Ellis, R.S. 1992, MNRAS, 254, 601 (BLE)

Bressan, A., Chiosi, C. \& Tantalo, R., 1996, A\&A, 311, 425

Bressan, A., Fagotto, F., Bertelli, G., \& Chiosi, C. 1993, A\&AS, 100, 647

Bruzual A., G., Barbuy, B., Ortolani, S., Bica, E., Cuisinier, F., \& Schiavon,
R. 1997, AJ, submitted

Carollo, C.M., de Zeeuw, P.T., van der Marel, R.P., Danziger, I.J. \& Qian,
E.E., 1995,
ApJ, 441, L25

Charlot, S., \& Silk, J. 1994, ApJ, 432, 453

Charlot, S., Worthey, G., \& Bressan, A. 1996, ApJ, 457, 625

Chiosi, C., Bressan, A., Portinaru, L., \& Tantalo, R. 1996, A\&A, submitted

Cole, S., Arag\'on-Salamanca, A., Frenk, C.S., Navarro, J.F.
\& Zepf, S.E., 1994, MNRAS, 271, 781

Davies, R.L., Sadler, E. \& Peletier, R., 1993, MNRAS, 262, 650

Dickinson, M., 1996, in eds Buzzoni, A., Renzini, A. \& Serrano, A., eds, Fresh
Views on Elliptical Galaxies,
ASP Conf. Series Vol. 86, p283

Djorgovski, S. \& Santiago, B.X., 1993, in Danziger, J. et al., eds,  Proc. of
the ESO/EIPC Workshop on Structure
Dynamics and Chemical Evolution of Elliptical Galaxies, ESO
Publication 45, p59

van Dokkum, P.G. \& Franx, M, 1996, MNRAS, 281, 985

Dressler, A., 1984, ApJ, 286, 97

Dressler, A., Lynden-Bell, D., Burstein, D., Davies, R.L., Faber, S.M.,
Terlevich, R.J. \&
Wegner, G., 1987, ApJ, 313, 42

Ellis, R.S., Smail, I., Dressler. A., Couch, W.J, 1996, preprint,
astro-ph/9607154

Faber, S.M., 1977, in Tinsley, B.M. \& Larson, R.B., eds,  The Evolution of
Galaxies and Stellar populations,
(Yale Observatory, New Haven), p157

Fagotto, F., Bressan, A., Bertelli, G., \& Chiosi, C. 1994a, A\&AS, 100, 647

---------. 1994b, A\&AS, 104, 365

---------. 1994c, A\&AS, 105, 29

Fluks, M. et al. 1994, A\&AS, 105, 311

Forman, W., Jones, C. \& Tucker, W., 1994, ApJ, 429, 77

Franx, M., van Gorkom, J. \& de Zeeuw, P.T., 1194, ApJ, 334, 613

Girardi, L., Bressan, A., Chiosi, C., Bertelli, G., \& Nasi, E. 1996,
A\&AS, 117, 113

Graham, A. \& Colless, M., 1997, MNRAS, submitted

Greggio, L. 1997, MNRAS 285, 151

Gunn, J.E., 1982, in Bruck, H.A., Coyne, G.V. \& Longair, M.S., eds,
Astrophysical Cosmology,
(Vatican City: Potificia Academia Scientarium), 233

Guzman, R., Lucey, J.R. \& Bower, R.G., 1993, MNRAS, 265, 731
Hjorth, J. \& Madsen, J., 1995, ApJ, 445, 55

Hui, X., Ford, H.C., Freeman, K.C. \& Dopita, M.A., 1995, ApJ, 449, 592

Iglesias, C.A., Rogers, F.J., \& Wilson, B.G. 1992, ApJ, 397, 717

Jones, L.A., \& Worthey, G. 1995, ApJ, 446, L31

J{\o}rgensen, I., Franx, M. \& Kjaergaard, P., 1996, MNRAS, 280, 167

Kauffmann, G., 1996a, MNRAS, 281, 475

Kauffmann, G., 1996b, MNRAS, 281, 487

Kauffmann, G. \& White, S.D.M. 1993, MNRAS, 261, 921

Kauffmann, G., White, S.D.M. \& Guiderdoni, B. 1993, MNRAS, 264, 201 (KWG)

Kauffmann, G., Charlot, S. \& White, S.D.M., 1996, MNRAS, 283, L117

Kennicutt, R.C. 1983, ApJ, 272, 54

Kodama, T. \& Arimoto, N., 1997, A\&A, in press

Kraan-Korteweg, R.C., 1981, A\&A, 104, 280

Kurucz, R.L. 1992, in IAU Symp. 149, The Stellar Populations of Galaxies, ed.
B.
Barbuy \& A. Renzini (Dordrecht: Kluwer), 225

Lacey, C., Guiderdono, B., Rocca-Volmerange, B. \& Silk, J., 1993, ApJ, 402, 15

Larson, R.B., 1974, MNRAS, 166, 585

Lejeune, T., Cuisinier, F., \& Buser, R. 1997a, A\&A, in press

Loewenstein, M. \& Mushotsky, R.F., 1996, ApJ, 466, 695

Maoz, D. \& Rix, H.-W., 1993, ApJ, 416, 215

Matteuci, F. \& Tornamb\'{e},F., 1987, A\&A, 185, 51

Mobasher, B., Guzman, R., Arag\'on-Salamanca, A. \& Zepf, S., 1997, MNRAS,
submitted

Mould, J.R., Oke, J.B., de Zeeuw, P.T. \& Nemec, J.M., 1990, AJ, 99, 1823

Mushotsky, R.F. \& Loewenstein, M., 1997, preprint, astro-ph/9702149

Renzini, A., Ciotti, L., Dercole, A. \& Pellegrini, S., 1993, ApJ, 419, 52

Rix, H.-W., de Zeeuw, P.T., Carollo, C.M., Cretton, N. \& van der Marel, M.
1997, preprint,
astro-ph/9702126

Roberts, M.S. \& Haynes, M.P., 1994, ARA\&A, 32, 115

Scalo, J.N., 1986, Fundamentals of Cosmic Physics, Vol. 11, p1

Stanford, S.A., Eisenhardt, P. \& Dickinson, M., 1995, ApJ, 450, 512

Tinsley, B.M., 1980, Fundamentals of Cosmic Physics, Vol. 5, p287

Tsujimoto, T., Nomoto, K., Yoshii, Y., Hashimoto, M., Yanagida, S. \&
Thielemann, F.-K., 1995,
MNRAS 277, 945

Spitzer, L., 1990, ARA\&A, 28, 71

Turner, M.S., 1996, preprint, astro-ph/9610158

Vader, J.P., 1986, ApJ, 305, 669

White, D.A. \& Fabian, A.C., 1995, MNRAS, 273, 72

White, S.D.M. \& Frenk, C.S., 1991, ApJ, 379, 52

White, S.D.M., Navarro, J.F., Evrard, A.E. \& Frenk, C.S., 1993, Nature, 366,
429

Worthey, G., 1994, ApJS, 95, 107

Worthey, G., Faber, S.M., \& Gonzalez, J.J. 1992, ApJ, 398, 69

Worthey, G., Faber, S.M., Gonzalez, J.J., Burstein, D. 1994, ApJS, 94, 687

Worthey, G., 1996, in eds Buzzoni, A., Renzini, A. \& Serrano, A., eds, Fresh
Views on Elliptical Galaxies,
ASP Conf. Series Vol. 86, p203

\end {document}